\documentstyle[12pt,thmsa,sw20lart]{article}
\input tcilatex
\QQQ{Language}{
American English
}

\begin{document}

\author{A. Tartaglia \\
%EndAName
Dip. Fisica, Politecnico,\\
Corso Duca degli Abruzzi 24, I-10129 Torino, Italy.\\
e-mail: TARTAGLIA @ITOPOLI.BITNET}
\title{FOUR DIMENSIONAL ELASTICITY AND GENERAL RELATIVITY }
\date{(To appear on GRAVITATION \& COSMOLOGY)}
\maketitle

\begin{abstract}
It has been shown that the extension of the elasticity theory in more than
three dimensions allows a description of space-time as a properly stressed
medium, even recovering the Minkowski metric in the case of uniaxial stress.
The fundamental equation for the metric in the theory is shown to be the
equilibrium equation for the medium. Examples of spherical and cylindrical
symmetries in four dimensions are considered, evidencing convergencies and
divergencies with the classical general relativity theory. Finally the
possible meaning of the dynamics of the four dimensional elastic medium is
discussed.
\end{abstract}

\section{Introduction}

The tensorial theory of elasticity in three dimensions has some apparent
similarity with the classical general relativity theory. The question I have
addressed in this paper is whether or not this formal analogy may correspond
to something more profound than the mere use of symmetric tensors in both
cases. In fact many authors have tried and introduced elasticity into
general relativity, casting the general equations into relativistic form\cite
{vari}\cite{rayner}\cite{sakharov}\cite{papapetrou}\cite{maugin}\cite{katz}.
This was usually made for ''practical'' purposes, in order to describe the
dynamic behaviour of astrophysical bodies in relativistic conditions, the
interaction of gravitational waves with bar antennas, the propagation of
shock waves in viscoelastic media and the like.

Here the approach is different, because the space-time itself will be looked
at as an elastic medium. Some hint in that direction can be found in the
literature, for instance in \cite{gerlach} where Gerlach and Scott introduce
a sort of ''elasticity of the metric'', though in connection with the
presence of matter.

The guiding idea of this paper is as follows: suppose space-time is a four
dimensional elastic medium. This, when unstrained, is perfectly homogeneous
and isotropic. The fundamental symmetry around any point inside it is
GL(4,R). Apply now some stresses to the medium: the symmetry will be broken
and reduced. In particular if the applied stress is one-dimensional the
consequence will be that one particular direction is specialized: could this
be time? Otherwise stated: is it possible that a uniaxial stress reduces the
GL(4,R) symmetry to SO(3,1)?

This approach, as we shall see, is indeed viable at a first level but leads
(I would say 'of course') to final equilibrium equations which are different
from those of general relativity. First of all the linear theory of
elasticity in any number of dimensions leads inevitably to linear equations.
The description of space-time that comes out as a result is ''static'' i.e.
perfectly deterministic. Any dynamics of a four-dimensional medium needs a
fifth evolution parameter and the evolution itself would bring about
modifications both in the past and the future of a given event.

In what follows I shall review the theory of elasticity and show how it can
be brought to describe a reasonable space-time.

\section{Instant review of the elasticity theory}

\subsection{\it Strain}

Suppose you have an n-dimensional elastic medium. In the absence of any
strain the geometry inside it is Euclidean (or at least, assume it is). The
squared distance between two nearby points is
\begin{equation}  \label{euclide}
dl_o^2=\epsilon _{\alpha \beta }dx^\alpha dx^\beta
\end{equation}

where $\epsilon _{\mu \nu \text{ }}$is the metric tensor of the unstrained
medium.

Introduce now an infinitesimal strain. Any point will in general be
displaced by a small vector $\overrightarrow{w}$, varying from place to
place. As a consequence the new squared distance between two points will be
written:
\begin{equation}  \label{strained}
dl^2=\left( \epsilon _{\alpha \beta }+d\epsilon _{\alpha \beta }\right)
\left( dx^\alpha +dw^\alpha \right) \left( dx^\beta +dw^\beta \right)
\end{equation}

The $x$'s are still referred to the unperturbed background and the $%
\overrightarrow{w}$'s as well as $\epsilon $'s are functions of the $x$'s.

In explicit form one has:
\begin{equation}  \label{variaz}
\left\{
\begin{array}{c}
\delta \epsilon _{\alpha \beta }=\epsilon _{\alpha \beta ,\mu }w^\mu \\
\delta w^\alpha =w_{,\mu }^\alpha dx^\mu
\end{array}
\right.
\end{equation}

Commas denote partial derivatives.

Now eq. (\ref{strained}) may be written:
\begin{equation}  \label{totale}
\begin{array}{c}
dl^2=dx^\alpha dx^\beta (\epsilon _{\alpha \beta }+\epsilon _{\alpha \mu
}w_{,\beta }^\mu +\epsilon _{\mu \beta }w_{,\alpha }^\mu +\epsilon _{\mu \nu
}w_{,\alpha }^\mu w_{,\beta }^\nu +\epsilon _{\alpha \beta ,\mu }w^\mu
+\epsilon _{\alpha \mu ,\nu }w^\nu w_{,\beta }^\mu + \\
+\epsilon _{\mu \beta ,\nu }w^\nu w_{,\alpha }^\mu +\epsilon _{\mu \nu
,\lambda }w^\lambda w_{,\alpha }^\mu w_{,\beta }^v)
\end{array}
\end{equation}

Usually part of the content of the brackets in (\ref{totale}) is identified
with the strain tensor $u_{\mu \nu }$, which is manifestly symmetric.

Now (\ref{totale}) becomes:
\begin{equation}  \label{ridotto}
dl^2=(\epsilon _{\alpha \beta }+u_{\alpha \beta })dx^\alpha dx^\beta
\end{equation}

The almost obvious identification:
\begin{equation}  \label{metrica}
g_{\mu \nu }\equiv \epsilon _{\mu \nu }+u_{\mu \nu }
\end{equation}

leads to
\begin{equation}  \label{linea}
dl^2=g_{\mu \nu }dx^\mu dx^\nu
\end{equation}

The symmetric tensor $g_{\mu \nu }$ is now the metric tensor of the strained
medium.

All this, as said, has been written using the unperturbed euclidean
coordinates. It is more natural to have recourse to internal or intrinsic
coordinates (those attached to the medium); these (let us call them $\xi $%
's) will in general be functions of the $x$'s. The geometric nature of the
objects used in the theory is such that it is possible to recast everything
in terms of the $\xi $'s by standard coordinate transformations for tensors.
In practice we can simply read formulas from (\ref{ridotto}) to (\ref{linea}%
) as if the $x$'s were $\xi $'s and nothing changes, consequently we shall
continue to use $x$'s in the new meaning\cite{hernandes}. By the way in the
base situation (unstrained medium) $x$'s and $\xi $'s coincide.

\subsection{\it Stress}

In the classical theory of elasticity a stress tensor is introduced whose
element $\sigma ^{\alpha \beta }$ has the meaning of the $\alpha $-component
of the force per unit surface acting on a surface element orthogonal to the $%
\beta $-direction. Assuming the range of the stresses to be zero
(propagation only by surface interaction), $\sigma ^{\mu \nu }$ is
symmetric. The next step is to link stresses and strains. This can be done
via the so called Hooke's law, which is indeed a linearization stating the
proportionality between stresses and strains. The theory may be found in any
text book on elasticity such as for instance \cite{landau}. The Hooke's law
in $n$ dimensions is expressed by the equivalent formulae:
\begin{equation}  \label{Hooke}
\left\{
\begin{array}{c}
\sigma _{\alpha \beta }=\left( K- \frac{2\mu }n\right) \epsilon _{\alpha
\beta }\epsilon ^{\nu \lambda }u_{\lambda \nu }+2\mu u_{\alpha \beta }\quad
\\
u_{\alpha \beta }=\left( \frac 1{n^2K}-\frac 1{2\mu n}\right) \epsilon
_{\alpha \beta }\epsilon ^{\nu \lambda }\sigma _{\nu \lambda }+\frac 1{2\mu
}\sigma _{\alpha \beta }
\end{array}
\right.
\end{equation}

$K$ is the uniform compression modulus; $\mu $ is the shear modulus.
Reasonable restrictions to the values of $K$ and $\mu $ are:
\begin{equation}  \label{vincoli}
K>0,\mu >0
\end{equation}

In general in a linearized theory of elasticity in any dimensions two
independent parameters are enough to describe the properties of the medium.
The parameters in use may be variously combined to produce others such as
the first Lam\'e coefficient $\lambda $ (the second is $\mu $), the Young
modulus $E$, the Poisson coefficient $\sigma $.

\section{Equilibrium conditions}

In our homogeneous stressed medium equilibrium is attained when the
following equation holds:
\begin{equation}  \label{equili}
\sigma _{\alpha \beta ,\beta }+f_\alpha =0
\end{equation}

Now $f_\alpha $ represents the $\alpha $-component of any force per unit
volume; considering the linearity of the Hooke's law indices are raised and
lowered using $\epsilon _{\mu \nu }$'s.

Combining (\ref{Hooke}), (\ref{equili}) and (\ref{metrica}) one obtains:
\begin{equation}  \label{equa}
\begin{array}{c}
\left[ \left( K- \frac{2\mu }n\right) \left( \epsilon ^{\nu \lambda
}g_{\lambda \nu }-n\right) -2\mu \right] \epsilon _{\alpha \beta ,\beta
}+\left( K-\frac{2\mu }n\right) \epsilon _{\alpha \beta }\left( \epsilon
^{\mu \nu }g_{\mu \nu }\right) _{,\beta }+ \\
+2\mu g_{\alpha \beta ,\beta }=-f_\alpha
\end{array}
\end{equation}

Using Cartesian coordinates (and Euclidean background geometry) (\ref{equa})
simplifies to
\begin{equation}  \label{base}
\left( K-\frac{2\mu }n\right) g_{\alpha \alpha ,\mu }+2\mu g_{\mu \nu ,\nu
}=-f_\mu
\end{equation}

Formulae (\ref{equa}) or (\ref{base}) are $n$ equations in $n(n+1)/2$
unknowns, consequently the problem is underdetermined. Suitable boundary
conditions are needed.

\section{Uniaxial stress.}

Let us now suppose that in our homogeneous $n$-dimensional medium a uniform
stress is applied along an arbitrary direction: let us call the
corresponding axis the $\tau $ axis. The stress tensor (as referred to a
Cartesian coordinates system) is, in our conditions:
\begin{equation}  \label{sforzi}
\left\{
\begin{array}{c}
\sigma _{oo}=p\qquad \qquad \quad \\
\sigma _{\alpha \beta }=0\qquad \alpha \neq \beta \\
\ \sigma _{ii}=\Sigma \qquad \quad \qquad
\end{array}
\right.
\end{equation}

The index number 0 corresponds to $\tau $, latin indices run from $1$ to $%
n-1 $. $\Sigma $ and $p$ are constants; $p>0$ means traction, $p<0$ means
compression.

Looking at eq.(\ref{base}) we see that any $g^{\mu \nu }$ =constant is a
solution of the equilibrium equation. To actually solve the problem we have
to directly deduce the $g^{\mu \nu }$ 's from (\ref{metrica}) and the
Hooke's law (\ref{Hooke}):
\begin{equation}  \label{defor}
\left\{
\begin{array}{c}
u_{oo}=\frac 1n\left[ \left( \frac 1{nK}+ \frac{n-1}{2\mu }\right) p+\left(
n-1\right) \left( \frac 1{nK}-\frac 1{2\mu }\right) \Sigma \right] \\
u_{\alpha \beta }=0\qquad \qquad \qquad \qquad \qquad \qquad \qquad \alpha
\neq \beta \\
u_{ii}=\frac 1n\left[ \left( \frac 1{nK}-\frac 1{2\mu }\right) p+\left(
\frac{n-1}{nK}+\frac 1{2\mu }\right) \Sigma \right] \qquad \qquad
\end{array}
\right.
\end{equation}

Now applying (\ref{metrica}) we see that it is $g_{\mu \nu }=\eta _{\mu \nu
} $, where $\eta _{\mu \nu }$ is the Minkowski metric tensor, whenever
\begin{equation}  \label{Minko}
\left\{
\begin{array}{c}
p= \frac{n-1}n\left( 2\mu -nK\right) \\
\Sigma =-\frac{2\mu +nK\left( n-1\right) }n\quad
\end{array}
\right.
\end{equation}

In four dimensions it is:
\begin{equation}  \label{quattro}
\left\{
\begin{array}{c}
p=\frac 32(\mu -2K)\quad \\
\Sigma =-\frac 12\left( \mu +6K\right)
\end{array}
\right.
\end{equation}

Considering conditions (\ref{vincoli}) it comes out that $\Sigma $$<0$ in
any case, which means transverse compression. This is consistent with what
we know from three-dimensional elasticity if $p>0$, i.e. if there is
traction along the $\tau $ axis. The parameter $p$ is actually greater than
zero when
\begin{equation}  \label{condi}
\mu >2K
\end{equation}

Minkowski space-time looks like a four-dimensional medium with suitable
elastic properties stretched along the time axis. Before the application of
the stress no difference exists among the various coordinates, so there is
no ''time''; once the stress is there one of the coordinates, measured along
any axis within the light cone about $\tau $, becomes no longer
interchangeable with the others: this, from the intrinsic view point is time.

\section{Spatially flat expanding universe.}

Another interesting case is that of an open expanding universe. The
corresponding conformally flat metric, in Cartesian coordinates, may be
written as:
\begin{equation}  \label{espandi}
ds^2=\alpha ^2\left( \tau \right) \left( d\tau ^2-dx^2-dy^2-dz^2\right)
\end{equation}

Introducing the metric (\ref{espandi}) into (\ref{base}) it is easily
verified that a nontrivial solution is found for $\alpha (\tau )$ if:
\[
f_i=0,\qquad f_o=F=\text{constant}
\]

The solution is:
\begin{equation}  \label{alpha}
\alpha ^2\left( \tau \right) =\frac F{2K-3\mu }\tau +\text{constant}
\end{equation}

and is consistent with the existence of a uniform volume field orthogonal to
any space section of the four-dimensional elastic medium.

Rescaling time according to the equation:
\[
\alpha \left( \tau \right) d\tau =dt
\]

the line element assumes its synchronous form:

\begin{equation}  \label{sincro}
ds^2=dt^2-a^2(t)(dx^2+dy^2+dz^2)
\end{equation}
with

\begin{equation}  \label{aquadrato}
a^2(t)=\left| \frac F{3\mu -2K}\right| ^{2/3}t^{2/3}
\end{equation}
As it can be seen the time dependence of the space scale factor is the same
as that for a matter dominated Friedmann universe \cite{gravi}.

The solution we have found corresponds to a spherically symmetric situation
in four dimensions. There is one center of symmetry (the big bang) and any
radial axis may be used as a time axis. Instead a more general positive or
negative space curvature Robertson Walker metric does not comply with this
symmetry and is no solution to eq. (\ref{equa}).

\section{Rotation symmetry about an axis.}

This is the typical situation which in general relativity leads, in the
static case, to the Schwarzschild solution. A general form for a metric with
this symmetry is:
\begin{equation}  \label{sferica}
g=\left(
\begin{array}{cccc}
\text{f}(r,\tau ) & 0 & 0 & 0 \\
0 & -h(r,\tau ) & 0 & 0 \\
0 & 0 & -r^2 & 0 \\
0 & 0 & 0 & -r^2\sin {}^2\vartheta
\end{array}
\right)
\end{equation}

Cylindrical coordinates $\tau ,r,\vartheta ,\varphi $ have been used. The
corresponding expression for the $\epsilon $'s is:
\begin{equation}  \label{epsi}
\epsilon =\left(
\begin{array}{cccc}
1 & 0 & 0 & 0 \\
0 & 1 & 0 & 0 \\
0 & 0 & r^2 & 0 \\
0 & 0 & 0 & r^2\sin {}^2\vartheta
\end{array}
\right)
\end{equation}

Inserting (\ref{sferica}) and (\ref{epsi}) into (\ref{equa}) leads to a
couple of independent equations:
\begin{equation}  \label{Schwa}
\left\{
\begin{array}{c}
\left( K-2\mu \right) \left( \text{\.f}-\dot h\right) +2\mu \text{\.f}=-f_o
\\
\left( K-2\mu \right) \left( \text{f}^{\prime }-h^{\prime }\right) -2\mu
\text{f}^{\prime }=-f_r
\end{array}
\right.
\end{equation}

Dots stand for partial derivatives with respect to $\tau $ and primes for
partial derivatives with respect to $r$.

The static case (independence of f and h from $\tau $) would require $f_o$
to be zero, whereas non trivial solutions exist only if $f_r\neq 0$. To
actually solve the problem one needs to impose the distribution of strains
or stresses on a suitable surface, remembering also that it should be f$,h>0$%
{}.

At the same results one can arrive starting from the solution of the
uniaxial stress case and letting $p$ and $\Sigma $ depend on $\tau $ and $r$.

\section{Discussion.}

It has been shown that the solutions for the equilibrium conditions inside a
four-dimensional elastic medium stressed in any way may provide reasonable
forms for the metric of space-time in various symmetry conditions. There are
however some problems: one is that of signature.

Treating the case of uniaxial stress we saw that it is possible to recover
the Minkowski metric. In four dimensions (\ref{defor}) and (\ref{quattro})
lead to the strain tensor components:
\begin{equation}  \label{strain}
\left\{
\begin{array}{c}
u_{oo}=0 \\
u_{ii}=-2
\end{array}
\right.
\end{equation}

However we know that the strain tensor is defined starting from a strain
vector field according to (\ref{totale}). In the case of uniaxial symmetry
the explicit form of the strain tensor in the background euclidean
coordinates is:

\begin{equation}  \label{tensore}
u_{\mu \nu }=w_{\mu ,\nu }+w_{\nu ,\mu }+w_{,\mu }^\alpha w_{\alpha ,\nu }
\end{equation}
Introducing (\ref{strain}) into (\ref{tensore}) and solving for the w's one
obtains:
\begin{equation}  \label{vettore}
\left\{
\begin{array}{c}
w^o= \text{constant}-2\tau \\
w^r=\left( -1\mp i\right) r\quad
\end{array}
\right.
\end{equation}

While the strain tensor is real the strain vector field is complex: this is
the price to be payed for the Minkowski signature.

Another important point to remind is that the theory, from Hooke's law on,
is linear. This implies that only weak field regions may be described this
way. It is possible to attain better approximations for stronger fields
having recourse to non linear elasticity. The starting point is the
development of the Helmholz free energy $F$ of the medium in powers of the
strains, where Hooke's law comes from. The next approximation after the
linear one is:
\begin{equation}  \label{Helmholtz}
F=F_o+\frac \lambda 2\left( u_\alpha ^\alpha \right) ^2+\mu u_{\alpha \mu
}u^{\mu \alpha }+\frac \nu 3\left( u_\alpha ^\alpha \right) ^3+\pi u_\alpha
^\alpha u_\mu ^\nu u_\nu ^\mu +\rho u_\alpha ^\beta u_\beta ^\mu u_\mu
^\alpha +...
\end{equation}

Three new parameters ($\nu ,\pi ,\rho $) have been introduced to
characterize the behaviour of the medium. Now it is no longer allowed to
raise and lower indices using simply the $\epsilon $'s; the $g_{\mu \nu }$'s
must be used after developing them up to first order in the $u$'s.
Everything is much more complicated but it may be managed.

Finally we may remark that our treatment of an equilibrium condition
corresponds to a perfectly static situation, i.e. to an entirely
deterministic universe where the histories coincide with the flux lines of
the strain vector field. However any elastic medium has not only statics but
also dynamics: it may vibrate and has characteristic internal frequencies.
If we consider a four dimensional elastic medium, vibrations have meaning of
course only with respect to some appropriate evolution parameter, let us
call it $T$: something like the good old newtonian time. Four-dimensional
observers have of course no means to measure $T$: their clocks actually
measure what we called $\tau $ (or t), though now $\tau $, as well as the
other space coordinates, parametrically depend on $T$ . An influence of the
vibrations in four plus one dimensions may however be seen from inside the
four-dimensional world.

Suppose for instance that in the medium there are a couple of points hold
fixed, whereas the rest undergoes elastic vibrations. Any strain flux line
(or history) going from one point to the other is continuously modified by
the vibrations in $T$. An internal observer, not being able to perceive $T$,
will notice that there are many nearby histories and what in four plus one
dimensions is a $T$ evolution for him may well be transformed into different
probabilities to be attached to the various histories. If the
four-dimensional observer wants to forecast future he will be led to average
over histories. A remarkable feature is the fact that for a given vibrating
point both future and past (in $\tau $ or $t$) vary in $T$.

I think that this view point may provide a new approach to quantum mechanics
and in any case is worth further investigation.

{}.

\end{document}